\documentclass[12pt]{article}
\usepackage{graphicx}
\begin{document}

\pagestyle{empty}

\noindent
{\bf A Study of Pulsation and Fadings in some R Coronae Borealis Stars}

\bigskip

\noindent
{\bf John R. Percy\\Department of Astronomy and Astrophysics, and\\Dunlap Institute of Astronomy and Astrophysics\\University of Toronto\\Toronto ON\\Canada M5S 3H4\\john.percy@utoronto.ca}

\medskip

\noindent
{\bf Kevin H. Dembski\\Department of Astronomy and Astrophysics\\University of Toronto\\Toronto ON\\Canada M5S 3H4\\kevin.dembski@mail.utoronto.ca}

\bigskip

{\bf Abstract}  We have measured the times of onset of recent fadings in four 
R CrB stars -- V854 Cen, RY Sgr, R CrB, and S Aps.  These times
continue to be locked to the stars' pulsation periods, though with some 
scatter.  In RY Sgr, the onsets of fading tend to occur at or a few days after
pulsation maximum.   We have studied the pulsation properties of RY Sgr through its recent long
maximum using (O-C) analysis and wavelet analysis.  
The period ``wanders" by a few percent.  This wandering can be
modelled by random cycle-to-cycle period fluctuations, as in some other types of 
pulsating stars.  The pulsation amplitude varies between 0.05 and 0.25 in
visual light, non-periodically but on a time scale of about 20 pulsation periods.

\medskip

\noindent
AAVSO keywords = AAVSO International Database; Photometry, visual; pulsating variables; R Coronae Borealis stars; period analysis; amplitude analysis

\medskip

\noindent
ADS keywords = stars; stars: late-type; techniques: photometric; methods: statistical; stars: variable; stars: oscillations

\medskip

\noindent
{\bf 1. Introduction}

\smallskip

R Coronae Borealis stars are rare carbon-rich, hydrogen-poor, highly-evolved yellow supergiants
which undergo fadings of up to 10 magnitudes, then slowly return to normal
(maximum) brightness; see Clayton (2012) for an excellent review.  
Most or all R CrB stars also undergo small-amplitude pulsations with
periods of a few weeks.  Although it was once considered that the fadings
were random, it is now known that, in at least some R CrB stars, the fadings
are locked to the pulsation period i.e. the onsets of the fadings occur at about
the same phase of the pulsation cycle (Pugach 1977, Lawson {\it et al.} 1992, Crause {\it et al.} 2007, hereinafter CLH).  This suggests a causal connection: the pulsation ejects a cloud of gas and
dust; when this cools, the carbon condenses into soot; if the cloud lies
between the observer and the star, the star appears to fade; it slowly
reappears as the cloud disperses.  This implies that the ejection is not
radially symmetric; a cloud is ejected, not a shell. 

Our interest in
these stars was sparked by a somewhat-accidental encounter with the R CrB
star Z UMi (Percy and Qiu 2018).  It had been misclassified as a Mira star,
which is what we had been studying at the time.  This star did not have a definitive
pulsation period, but we measured the times of onset of its fadings, and we found that they were ``locked" to a period of 41.98 days, a typical pulsation period
for an R CrB star.

\medskip

\noindent
{\bf 2. Data and Analysis}

\smallskip

We used visual observations from the AAVSO International Database (AID:
Kafka 2018), the AAVSO VSTAR time-series analysis package (Benn 2013)
which includes Fourier analysis, wavelet analysis, and polynomial fitting routines,
and (O-C) analysis to study the five R CrB stars previously studied by
CLH, and to study the pulsation of RY Sgr in more detail.

\medskip

\noindent
{\bf 3 Results}

\medskip

\noindent
3.1. Pulsation-fading relationships in R CrB stars

\smallskip

CLH showed that, in five R CrB stars, the times of onset of
fadings were locked to their pulsation periods.  The five stars, and their
pulsation periods in days, were: V854 Cen (43.25), RY Sgr (37.79), UW Cen
(42.79), R CrB (42.97), and S Aps (42.99).

In the first part of this project, we determined the times of onset of fadings
of these five stars since the work of CLH.  
The determination of these times was non-trivial.  The visual observations had a
typical uncertainty of 0.2 magnitude.  Sometimes the data were sparse, and
the exact time of onset was not well covered by the observations.  This is
especially true if the onsets fell within the seasonal gaps in the stars'
observations.  Some onsets could therefore not be measured.  To determine
the times, we experimented with fitting horizontal lines to the
light curves preceding fadings, and sloping lines to the light curves
following the onset of fadings, as well as using ``by-eye" judgement.

We first
determined, independently, the times published by CLH.
Our times differed on average by $\pm$4 days, which is the typical uncertainty
of our determinations and CLH's.  The differences averaged only $\pm$3 days for
R CrB, presumably the most densely-observed star.  On average, our times were
+1 day later than CLH's, which is not significantly different from
zero.  Our times are given in Table 1. The ephemerides are the same as
used by CLH.  UW Cen
did not have any recent fadings whose times could be determined.
The first time listed for each star is our redetermination of the time of onset
of the last fading observed by CLH.  This is followed by the CLH
determination.  This provides an indication of the difference and uncertainty in the
timings.  Times are labelled with a colon (:) if there was some scatter and/or sparseness in the data, or with a double colon (::) if there
was much scatter and/or sparseness.

Figure 1 shows the light curves, using the same format as CLH.
The times of onset of fadings are indicated with an x.  Figure 2 shows
the (O-C)'s between our times of onset of fadings, and the pulsation
ephemerides used by CLH.  The average (O-C) for our times is almost
twice that for CLH's times.  This will be discussed in Section 4.

\begin{figure}
\begin{center}
\includegraphics[height=10cm]{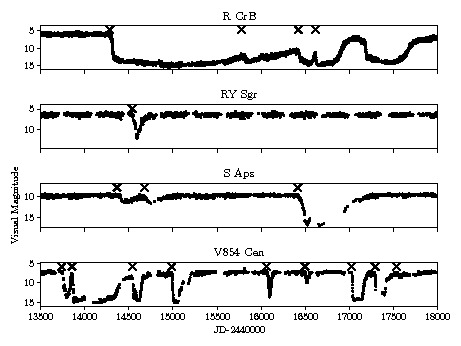}
\caption{The recent AAVSO visual light curves of R CrB, RY Sgr, S Aps,
and V854 Cen.  The times of fadings (Table 1), as measured by us, are marked with an x.}
\end{center}
\end{figure}

\begin{figure}
\begin{center}
\includegraphics[height=8cm]{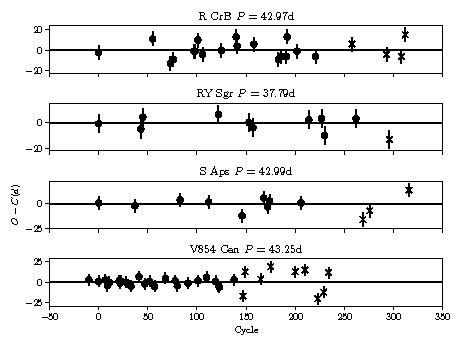}
\caption{The (O-C) diagrams for the times of onsets of fadings in R CrB, RY Sgr, S Aps,
and V854 Cen, using the periods given in section 3.1, and the observed times
and cycle numbers listed in Table 1.  The filled circles are the (O-C)'s
published by CLH.}
\end{center}
\end{figure}

\medskip

\noindent
3.2. Times of pulsation maximum in RY Sgr

\smallskip

RY Sgr has the largest pulsation amplitude of any known R CrB star, though
it is only about 0.15 in V.  Several groups have observed or discussed the pulsation of
RY Sgr for the purpose of determining and interpreting its apparent period
change: Kilkenny (1982), Lawson and Cottrell (1990), Lombard and Koen (1993),
Menzies and Feast (1997), among others.

RY Sgr has been at maximum since JD 2454900.  In order to investigate the
pulsation period, we have determined the times
of 58 
pulsation maxima between JD 2455031 and JD 2458243.  They are listed in Table 2.
They were determined independently by both of us, using low-order polynomial
fitting (KHD, JRP) and phase-curve fitting (JRP), and then appropriately averaged.

\smallskip

\medskip

\noindent
3.3. Pulsation period variations in RY Sgr

\smallskip

The authors who were mentioned in section 3.2 determined the apparent period change in
RY Sgr, and suggested various interpretations, including smoothly-varying
period changes, and abrupt period changes.  We have used two methods to
investigate the period change: (O-C) analysis, and wavelet analysis, and
applied them to the times in Table 2.

Figure 3 shows the (O-C) diagram for RY Sgr, using the times of maximum
listed in Table 2, and a period of 37.91 days.  The scatter is consistent
with the uncertainties in the times of maximum.  Figure 4 (top) shows the
period variation determined by wavelet analysis, using the WWZ routine in
VSTAR.  Both figures show that the period ``wanders" between values of 37.0 and 38.5
days, with the period being approximately constant in the first third of
the interval, increasing to a higher value in the second third, and decreasing
to a lower value in the final third.  The variation is not periodic, but its time scale is about 20 pulsation periods.  In Mira stars, the time scale averages about 40 pulsation periods (Percy and Qiu 2018).

\begin{figure}
\begin{center}
\includegraphics[height=7cm]{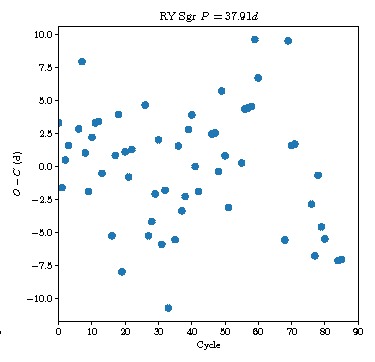}
\caption{The (O-C) diagram for the times of pulsation maximum of RY Sgr listed in
Table 2, using a period of 37.91 days.  The period is approximately constant
through cycles 0-25, slightly larger than average (upward slope) through cycles 25-60, and
slightly smaller than average (downward slope) through cycles 60-90.}
\end{center}
\end{figure}

\begin{figure}
\begin{center}
\includegraphics[height=7cm]{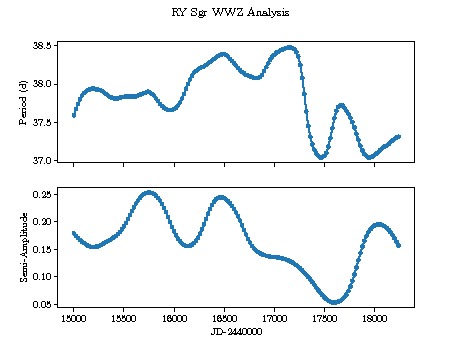}
\caption{The variation in the pulsation period (top) and amplitude (bottom) of RY Sgr versus
time, determined using the WWZ wavelet routine in VSTAR, and AAVSO visual observations.}
\end{center}
\end{figure}

\medskip

\noindent
3.4. Are the period variations due to random cycle-to-cycle fluctuations?

\smallskip

The ``wandering" pulsation periods of large-amplitude pulsating red giant
stars (Mira stars) have been modelled by random cycle-to-cycle period
fluctuations (Eddington and Plakidis 1929, Percy and Colivas 1999).  We
have investigated whether the period variations in RY Sgr can be modelled
in this way by applying the Eddington-Plakidis formalism to the times of
pulsation maximum given in Table 2.  For this, we used a program written
by one of us (KHD) in Python.  We first tested it (successfully) on times of maximum of
Mira, for comparison with Figure 1 in Percy and Colivas (1999).

In Figure 3, we showed the (O-C) values for RY Sgr, using a period of 37.91 days.
Then, following Eddington and Plakidis (1929): let
a(r) be the (O-C) of the rth maximum, and let ux(r) = a(r+x) - a(r), and
$\overline{ux}$ be the average value, without regard to sign, of ux(r) for as many
values of r as the observational material admits, 
then $\overline{ux}^2$ = $2a^2 + xe^2$
where a is the average observational error in determining
the time of maximum, and e the average fluctuation in period, per cycle.
A graph of $\overline{ux}^2$ versus x (the ``Eddington-Plakidis diagram") should be a straight line if random cycle-to-cycle fluctuations occur. 
Figure 5 shows the $\overline{ux}^2$ versus x graph for RY Sgr, using the
times listed in Table 2.  The graph is approximately linear, with scatter
which is not unexpected, given the limitations of the data.  The value  of 
a = 2.7 days is consistent with the errors in the measured times of maximum.

We also generated a $\overline{ux}^2$ versus x graph for RY Sgr, using the
times of pulsation maximum published by Lawson and Cottrell (1990).
They extend from JD 2441753 to 2447642.  The graph is shown in Figure 6.
The graph is approximately linear.
The slope is comparable with that
in Figure 5, and the value of a = 3.1 days is consistent with the expected
errors in the measured times of maximum.  The slopes e are 0.9 and 1.0 day
for our data and Lawson and Cottrell's, respectively.

\begin{figure}
\begin{center}
\includegraphics[height=7cm]{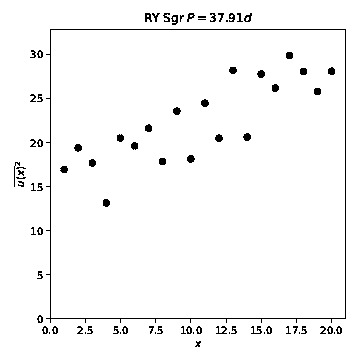}
\caption{The Eddington-Plakidis diagram for RY Sgr, using the times of pulsation
maximum in Table 2, and a period of 37.91 days.}
\end{center}
\end{figure}

\begin{figure}
\begin{center}
\includegraphics[height=7cm]{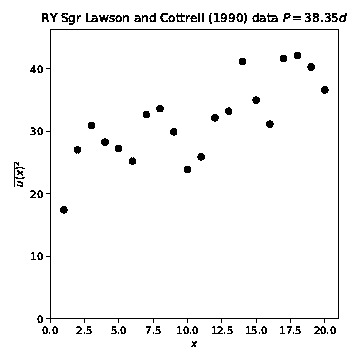}
\caption{The Eddington-Plakidis diagram for RY Sgr, using the times of pulsation
maximum given by Lawson and Cottrell (1990) between JD 2441753 and 2447642,
and a period of 38.56 days.}
\end{center}
\end{figure}

\medskip

\noindent
3.5. Pulsation amplitude variations in RY Sgr

\smallskip

Most of the famous Cepheid pulsating variables have constant pulsation 
amplitudes, but this is not true of other types, especially low-gravity
stars: pulsating red giants
(Percy and Abachi 2013), pulsating red supergiants (Percy and Khatu 2014)
and some pulsating yellow supergiants (Percy and Kim 2014).  The amplitudes
of these stars vary by up to a factor of ten, on time scales of
20-30 pulsation periods.

We have used wavelet analysis to determine the variation in pulsation amplitude
in RY Sgr, during the last 3000 days when the star was at
maximum (JD 2455000 to JD 2458250).  The results
are shown in the lower panel in Figure 4.  The visual amplitude varies between
0.05 and 0.25.  The amplitude variations can be confirmed by Fourier analysis
of subsets of the data.  The variation is not periodic, but occurs on a time scale
of about 20 pulsation periods.  This time scale is comparable to that found
in the pulsating star types mentioned above.  There is no strong consistency
to the direction of the changes, though there is a slight tendency for the
amplitude to be relatively medium-to-high at the beginning of a maximum, and relatively  medium-to-low
at the end.

We used the same method to study the amplitude variations during several
shorter intervals when the star was at maximum.  The results are given
in Table 3.  During these intervals, the visual amplitude also varies between
0.05 and 0.25.  The median time scale of visual amplitude variation is about 30 (range 15 to 55) 
pulsation periods.

\medskip

\noindent
3.6. At What Pulsation Phase Does the Onset of Fading Occur?

\smallskip

CLH stated ``... the absolute phase of the decline onsets could not be
determined from the AAVSO data ...".  We have attempted to make this
determination for RY Sgr, as follows.  For each of the times of onset of fading
determined by CLH, we have examined the previous 50-60 days of data,
and measured the times of pulsation maximum using the same methods as
in section 3.2.  The results are listed in Table 4.  The times of onset of
fadings are the predicted times, given by CLH.  There is considerable
scatter, as there was in measuring the times of maximum in Table 2. 
Before some fadings, the data were too sparse to measure the
pulsation maximum.
On average, the onsets of fading occur 7 days after pulsation
maximum.  According to Pugach (1977), the onset of fadings occurs at
pulsation maximum, and the same is true for V854 Cen (Lawson {\it et al.} 1992).
These conclusions depend, to some extent, on the definition of when the
onset occurs, and may not be in conflict.

\medskip

\noindent
{\bf 4. Discussion}

\smallskip

The times of onset of fadings that we have measured (Table 1, Figure 1) seem to continue
to be locked to the pulsation periods (Figure 2), though with a scatter $\pm$11 days which is twice that obtained by CLH, even though our measured times are consistent with theirs.  There are several possible explanations: (1) our times
are actually less accurate than theirs; (2) the ``wandering" period (Figures 3 and 4) causes
some scatter; (3) the fadings are not exactly locked to the pulsation;
there are random factors in the pulsation, mass ejection, and onset of fadings which add to the scatter; (4) the differences are a statistical anomaly.

We recognize that it is challenging to determine times of onset of
fadings, or times of pulsation maxima, using visual data.  This is
where the density of the visual data can often help.  

Our results (the linearity of Figures 5 and 6) suggest that the period 
variations can be modelled, at least in part, by random
cycle-to-cycle variations, as in Mira stars, rather than solely by a smooth
evolutionary variation, or an abrupt variation.  We cannot rule out
the presence of a small smooth or abrupt variation, but it is buried in the random
period-fluctuation noise.  The cause of the fluctuations is not known, but
may be connected with the presence of large convective cells in the outer
layers of the stars.  The fact that the star ejects clouds, rather than shells,
suggests that the outer layers of the star are not radially symmetric.

The discovery of a variable pulsation amplitude in RY Sgr (Figure 4) is an
interesting but not-unexpected result, given the presence of amplitude
variations in other low-gravity pulsating stars.  We note that, when
the pulsation amplitude is at its lowest, it is even more difficult to
measure the times of pulsation maximum.

There are other R CrB stars which are known or suspected to pulsate.
Rao and Lambert (2015) list 29.  Most if not all of them have pulsation
amplitudes which are even smaller than that of RY Sgr, so it will be almost
impossible to study their pulsation with visual data.  At least one of the 29 stars has an
incorrect period: Z UMi is listed as having a period of 130 days, but
Percy and Qiu (2018) were not able to fit AAVSO data to that period but,
as mentioned in the Introduction,
using the observed times of onset of fadings, they
suggested a period of 41.98 days instead.  It may be possible to
determine times of onset of fadings of some of the 29 stars, and see if
there is a period that they are locked to, as Percy and Qiu (2018) did for Z UMi.

In future, most of
these stars will be monitored through facilities such as LSST (the
Large Synoptic Survey Telescope), but
the century of archival AAVSO data will remain unique.

\medskip

\noindent
{\bf 5. Conclusions}

\smallskip

We have derived new information about the pulsation of the R CrB star RY Sgr,
especially about the variation of its period and amplitude.  We have also
strengthened the connection between the pulsation and the fadings in this
star.  We have used long-term archival visual data but, since the pulsation
amplitudes of other R CrB stars are even smaller than that of RY Sgr, future
studies like ours will have to use long-term precision photoelectric or
CCD observations.

\medskip

\noindent
{\bf Acknowledgements}

\smallskip

We thank the AAVSO observers who made the observations on which this project
is based, the AAVSO staff who archived them and made them publicly available, and the developers
of the VSTAR package which we used for analysis.  Coauthor KHD was a participant
in the University of Toronto Work-Study Program, which we thank for 
administrative and financial support.
This project made
use of the SIMBAD database, maintained in Strasbourg, France.
The Dunlap Institute is funded through an endowment established by the David Dunlap family and the University of Toronto.

\bigskip

\noindent
{\bf References}

\smallskip

\noindent
Benn, D. 2013, VSTAR data analysis software {http://www.aavso.org/vstar-overview).

\noindent
Crause, L.A., Lawson, W.A., and Henden, A.A. 2007, {\it Mon. Not. Roy. Astron. Soc.}, {\bf 375}, 301.

\noindent
Eddington, A.S., and Plakidis, S. 1929, {\it Mon. Not. Roy. Astron. Soc.}, {\bf 90}, 65.

\noindent
Kafka, S. 2018, variable star observations from the AAVSO International Database

(https://www.aavso.org/aavso-international-database)

\noindent
Kilkenny, D. 1982, {\it Mon. Not. Roy. Astron. Soc.}, {\bf 200}, 1019.

\noindent
Lawson, W.A., and Cottrell, P.L. 1990, {\it Mon. Not. Roy. Astron. Soc.},
{\bf 242}, 259.

\noindent
Lawson, W.A., {\it et al.} 1992, {\it Mon. Not. Roy. Astron. Soc.}, {\bf 256}, 339.

\noindent
Lombard, F., and Koen, C. 1993, {\it Mon. Not. Roy. Astron. Soc.}, {\bf 263}, 309.

\noindent
Menzies, J.W., and Feast, M.W. 1997, {\it Mon. Not. Roy. Astron. Soc.}, {\bf 285}, 358.

\noindent
Percy, J.R., and Colivas, T. 1999, {\it Publ. Astron. Soc. Pacific}, {\bf 111}, 94.

\noindent
Percy, J.R., and Abachi, R. 2013, {\it J. Amer. Assoc. Var. Star Obs.}, {\bf 41}, 193.

\noindent
Percy, J.R., and Khatu, V.C. 2014, {\it J. Amer. Assoc. Var. Star Obs.}, {\bf 42}, 1.

\noindent
Percy, J.R., and Kim, R.Y.H. 2014, {\it J. Amer. Assoc. Var. Star Obs.}, {
bf 42}, 267.

\noindent
Percy, J.R., and Qiu, A.L. 2018, arxiv.org/abs/1805.11027

\noindent
Pugach, A.F. 1977, {\it Inf. Bull. Var. Stars}, No. 1277.

\noindent
Rao, N.K., and Lambert, D.L. 2015, {\it Mon. Not. Roy. Astron. Soc.}, {\bf 447}, 3664.

\medskip

\medskip

\smallskip
\begin{table}
\begin{center}
\caption{New Times of Onset of Fadings in Four R CrB Stars}
\begin{tabular}{rclrrr}
\hline
Star & Cycle (n) & JD (obs) & JD (calc) & O-C (d) & Note \\
\hline
S Aps & 206 & 2451670 & 2451674 & -4 & PD \\
-- & 206 & 2451675 & 2451674 & 1 & CLH \\
-- & 269 & 2454366:: & 2454382 & -16 & -- \\
-- & 276 & 2454676: & 2454683 & -7 & -- \\
-- & 316 & 2456417 & 2456403 & 14 & -- \\
RY Sgr & 262 & 2453273 & 2453266 & 7 & PD \\
-- & 262 & 2453269 & 2453266 & 3 & CLH \\
-- & 296 & 2454538: & 2454551 & -13 & -- \\
V854 Cen & 138 & 2453376 & 2453368 & 8 & PD \\
-- & 138 & 2453371 & 2453368 & 3 & CLH \\
-- & 147 & 2453740:: & 2453757 & -17 & -- \\
-- & 149 & 2453856 & 2453843 & 13 & -- \\
-- & 165 & 2454540: & 2454536 & 4 & -- \\
-- & 175 & 2454987 & 2454968 & 19 & -- \\
-- & 200 & 2456062 & 2456049 & 13 & -- \\
-- & 210 & 2456497 & 2456482 & 15 & -- \\
-- & 223 & 2457024 & 2457044 & -20 & -- \\
-- & 229 & 2457292: & 2457304 & -12 & -- \\
-- & 234 & 2457532 & 2457520 & 12 & -- \\
R CrB & 221 & 2452678 & 2452689 & -11 & PD \\
-- & 221 & 2452683 & 2452689 & -6 & CLH \\
-- & 258 & 2454285 & 2454279 & 6 & -- \\
-- & 293 & 2455779 & 2455783 & -4 & -- \\
-- & 308 & 2456421 & 2456427 & -6 & -- \\
-- & 312 & 2456615 & 2456599 & 15 & -- \\
\hline
\end{tabular}
\end{center}
\end{table}

\smallskip
\begin{table}
\begin{center}
\caption{Times of Pulsation Maximum in RY Sgr (JD - 2400000)}
\begin{tabular}{cccc}
\hline
JD (max) & JD (max) & JD (max) & JD (max) \\
\hline
55031 & 55740 & 56466 & 57274 \\ 
55064 & 55787 & 56509 & 57309 \\ 
55104 & 55823 & 56548 & 57600 \\ 
55143 & 55863 & 56582 & 57653 \\ 
55258 & 56018 & 56618 & 57683 \\ 
55301 & 56046 & 56774 & 57721 \\ 
55332 & 56085 & 56812 & 57906 \\ 
55367 & 56125 & 56847 & 57940 \\ 
55409 & 56167 & 56891 & 57984 \\
55448 & 56197 & 56924 & 58018 \\
55486 & 56239 & 56958 & 58055 \\
55520 & 56268 & 57113 & 58205 \\ 
55629 & 56349 & 57155 & 58243 \\
55673 & 56394 & 57193 & -- \\
55714 & 56427 & 57231 & -- \\
\hline
\end{tabular}
\end{center}
\end{table}

\smallskip

\begin{table}
\begin{center}
\caption{Pulsation Amplitude Variations in RY Sgr}
\begin{tabular}{cc}
\hline
JD range & amplitude range \\
\hline
2432950-2435437 & 0.06-0.20 \\
2436597-2438054 & 0.14-0.20 \\
2438303-2439653 & 0.07-0.23 \\
2442095-2443300 & 0.17-0.11 \\
2445739-2447949 & 0.07-0.20 \\
2449886-2451403 & 0.10-0.24 \\
2452078-2453227 & 0.04-0.15 \\
\hline
\end{tabular}
\end{center}
\end{table}

\begin{table}
\begin{center}
\caption{Times of Onset of Fading and of Pulsation Maximum in RY Sgr}
\begin{tabular}{ccc}
\hline
Onset of Fading F & Pulsation Maximum M & F-M (d) \\
\hline
2443366 & 2443357 & +9 \\
2444990 & 2444970 & +20 \\
2445066 & 2445064 & +2 \\
2447976 & 2447977 & -1 \\
2449147 & 2449133 & +14 \\
2441452 & 2441458 & -6 \\
2452057 & 2452046 & +11 \\
2453266 & 2453260 & +6 \\
\hline
\end{tabular}
\end{center}
\end{table}

\end{document}